\documentclass[aps,prx,floatfix,twocolumn,superscriptaddress]{revtex4-1}

\usepackage{graphicx}
\usepackage[caption=false]{subfig}
\usepackage{natbib}
\usepackage{color}

\usepackage{nopageno, tikz, amsmath, amsfonts, mathrsfs,bm}

\newcommand{\fq}[1]{\Phi_\mathrm{ext}={#1}\Phi_0}
\newcommand{\fqn}{\Phi_\mathrm{ext}}
\newcommand{\g}{|g\rangle}
\newcommand{\e}{|e\rangle}

\begin{document}

\title{Simultaneous monitoring of fluxonium qubits in a waveguide}

\author{A.~Kou}
\affiliation{Departments of Applied Physics and Physics, Yale University, New Haven, CT 06520, USA}
\author{W.~C.~Smith}
\affiliation{Departments of Applied Physics and Physics, Yale University, New Haven, CT 06520, USA}
\author{U.~Vool}
\affiliation{Departments of Applied Physics and Physics, Yale University, New Haven, CT 06520, USA}
\author{I.~M.~Pop}
\affiliation{Departments of Applied Physics and Physics, Yale University, New Haven, CT 06520, USA}
\affiliation{Physikalisches Institut, Karlsruhe Institute of Technology, Karlsruhe 76131, Germany}
\author{K.~M.~Sliwa}
\affiliation{Departments of Applied Physics and Physics, Yale University, New Haven, CT 06520, USA}
\author{M.~Hatridge}
\affiliation{Departments of Applied Physics and Physics, Yale University, New Haven, CT 06520, USA}
\affiliation{Department of Physics, University of Pittsburgh, Pittsburgh, PA 15260, USA}
\author{L.~Frunzio}
\affiliation{Departments of Applied Physics and Physics, Yale University, New Haven, CT 06520, USA}
\author{M.~H.~Devoret}
\affiliation{Departments of Applied Physics and Physics, Yale University, New Haven, CT 06520, USA}

\date{\today}

\begin{abstract}
Most quantum-error correcting codes assume that the decoherence of each physical qubit is independent of the decoherence of any other physical qubit. We can test the validity of this assumption in an experimental setup where a microwave feedline couples to multiple qubits by examining correlations between the qubits. Here, we investigate the correlations between fluxonium qubits located in a single waveguide. Despite being in a wide-bandwidth electromagnetic environment, the qubits have measured relaxation times in excess of 100~$\mu$s. We use cascaded Josephson parametric amplifiers to measure the quantum jumps of two fluxonium qubits simultaneously. No correlations are observed between the relaxation times of the two fluxonium qubits, which indicates that the sources of relaxation are local to each qubit in our setup. Our correlation analysis can be generalized to different types of qubits and our architecture can easily be scaled to monitor larger numbers of qubits.

\end{abstract}


\maketitle
\section{Introduction}
Quantum hardware, which depends on superpositions of fragile quantum states, is much more susceptible to errors than classical hardware. In principle, however, by using quantum error correction, one can still reliably perform arbitrarily long quantum computations with faulty hardware provided that the error rate of the hardware is sufficiently small \cite{Aharonov1999}. The criteria for sufficiently small depends strongly on the type of noise coupled to the quantum hardware. If one assumes that the noise couples to each qubit independently, i.e. the error in each qubit is independent of the errors in all other qubits, then it was theoretically proven that arbitrarily long quantum computations could be performed with qubits that have error probabilities less than $10^{-5}$ \cite{Aliferis2005}. 

Standard error-correcting codes, however, offer poor protection against correlated errors. For quantum hardware with two-qubit correlations that decay algebraically with distance between the qubits, the error probability must be less than $10^{-10}$ when standard quantum error-correcting codes are used \cite{Aharonov2006}. For hardware with correlated errors, one should use alternative methods such as dynamical decoupling or decoherence-free subspaces with quantum error-correcting codes in order to accurately perform arbitrarily long quantum computations \cite{Lidar1998,Viola1999,Novais2006}. Hence, to properly perform quantum error correction, one must understand the correlations present in any proposed physical implementation of a quantum computer.

\begin{figure}[!t]
\includegraphics[width=86mm]{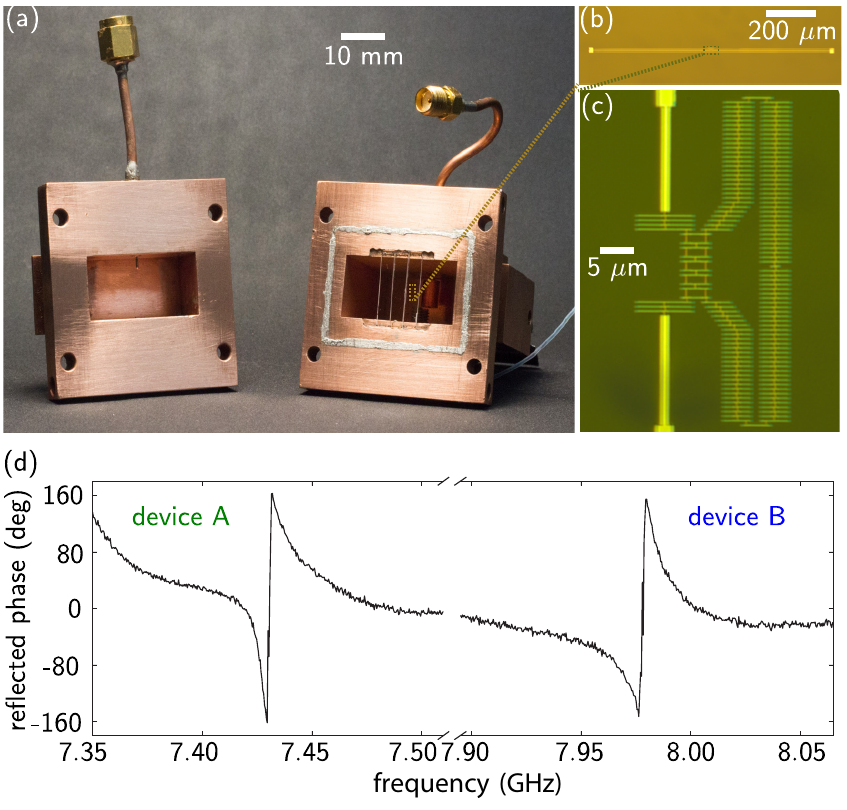}
\caption{(a) Waveguide used for frequency-multiplexed readout of fluxonium devices. Input signals were sent into the waveguide via the 50$\Omega$ impedance-matched coupler shown on the right. An input pin, shown on the left and located 3~mm away from the samples, coupled qubit drive tones to the devices. The devices were fabricated on sapphire chips, which were then placed inside of the waveguide. Superconducting wire wound around the waveguide provided an external magnetic flux bias. (b) Optical image of the antenna used to read out device A (shown in (c)). (d) Phase of the reflected signal from the waveguide as a function of drive frequency. The resonance associated with the readout antenna for device A and device B was observed at $7.430~\text{GHz}$ and $7.979~\text{GHz}$, respectively. \label{fig:waveguide}}
\end{figure}

Superconducting circuits have emerged as a promising platform for building quantum computers. Qubits based on superconducting circuits currently achieve coherence times on the order of $100~\mu$s with a typical gate time of 10~ns, which, assuming that the gate fidelities are limited by  coherence times, is an error probability per qubit of $10^{-4}$ \cite{Paik2011, Rigetti2012}. Quantum error-correcting codes have been demonstrated with small numbers of superconducting circuits \cite{Reed2012, Kelly2015, Riste2015, Ofek2016}. As these systems begin to scale up and approach thresholds required for error correction, it is vital to determine if there are correlated error channels in superconducting qubits. Previous measurements of multiple superconducting qubits have focused on crosstalk between qubits during gate operation and readout \cite{McDermott2005, Chen2012, Schmitt2014}. 

Here, we present the first real-time measurements of correlations between the relaxation rates of two superconducting qubits. Our experiment is based on using a novel low-loss waveguide for multiplexed readout of fluxonium qubits  \cite{Manucharyan2009}. Fluxonium qubits dispersively coupled to on-chip resonators have recently been demonstrated to have long relaxation times \cite{Pop2014,Vool2014} and can be easily incorporated into a multiplexed readout setup. We monitored the quantum jumps of each qubit simultaneously and examined temporal correlations between the relaxation rates of the fluxonium qubits. The observation of correlations would be indicative of the two qubits coupling to a changing common environment, which can be caused by a fluctuating density of background quasiparticles \cite{Pop2014,Vool2014} or stray electromagnetic fields. We found no correlations between the relaxation rates of the qubits up to the detection efficiency of our measurement setup. We conclude that the sources of relaxation in the qubits are local and discuss prospects for extending this measurement to larger numbers of qubits and finer resolution of correlations.
\section{Waveguide Implementation}
A WR-102 (with transverse inner dimensions of 1.020 in by 0.510 in) waveguide served as a low-loss wide-bandwidth electromagnetic environment for frequency multiplexed readout as shown in Fig.~\ref{fig:waveguide}(a). The waveguide was made with oxygen-free high-conductivity (OFHC) copper. Input signals from coaxial cables were coupled into the waveguide via OFHC 50$\Omega$ impedance-matched adapters. The insertion loss and bandwidth of the waveguide were adjusted with aluminum tuning screws. An indium seal ensured continuous electrical connection between the two ends of the waveguide. At room temperature, the waveguide had an insertion loss of $-0.3$ dB over a $6-8$~GHz band. 

Dipole antennae directly coupled to the lowest-order propagating electromagnetic mode of the waveguide were used to readout the fluxonium devices. The antennae were LC oscillators where the inductance was provided by Josephson junctions and the capacitance was provided by the long metal electrodes, as shown in Fig.~\ref{fig:waveguide}(b). The junctions were fabricated with Al/AlOx/Al using the bridge-free double-angle evaporation technique \cite{Lecocq2011}. Superconducting quantum interference devices served as tunable inductances to adjust the resonant frequencies of the antennae \cite{SQUIDnote}. The zero-field inductance of the antenna coupled to device A (B) was 22 nH (20 nH). Shared Josephson junctions inductively coupled each readout antenna to a fluxonium device. The shared inductance between the antennae and the fluxonium devices was the same for both devices -- 8.35 nH.

Each fluxonium device was composed of a small Josephson junction, which provides nonlinearity, in parallel with an array of 131 larger junctions as shown in Fig. \ref{fig:waveguide}(c). Each array had an inductance of 455 nH and served as the superinductance for each device. Both the devices and the antennae were fabricated on sapphire chips. The chips were placed in the waveguide a quarter-wavelength away from a copper wall, which situates them at an electric field antinode. An input pin located 3~mm away from the chips coupled qubit drive tones to the two devices when the ground-excited state transition frequency of the devices was below the lower cutoff frequency of the waveguide.

The waveguide was housed in an aluminum shield coated with infrared-absorbing material to protect against infrared radiation and offset magnetic fields. A $\mu-$metal shield enclosing the aluminum shield further screened stray magnetic fields \cite{KurtisThesis}. The waveguide and shields were thermalized to the mixing chamber plate of a dilution refrigerator with a base temperature of $\sim20$~mK. 

The phase of the reflected signal from the waveguide is shown in Fig.~\ref{fig:waveguide}(d). We observed resonances associated with the readout antennae for device A and device B. The readout antenna for device A (B) has a resonant frequency of 7.430~GHz (7.979~GHZ) and a linewidth of $\kappa/2 \pi=10$~MHz (14~MHz).

\begin{figure}[!t]
\includegraphics[width=86mm]{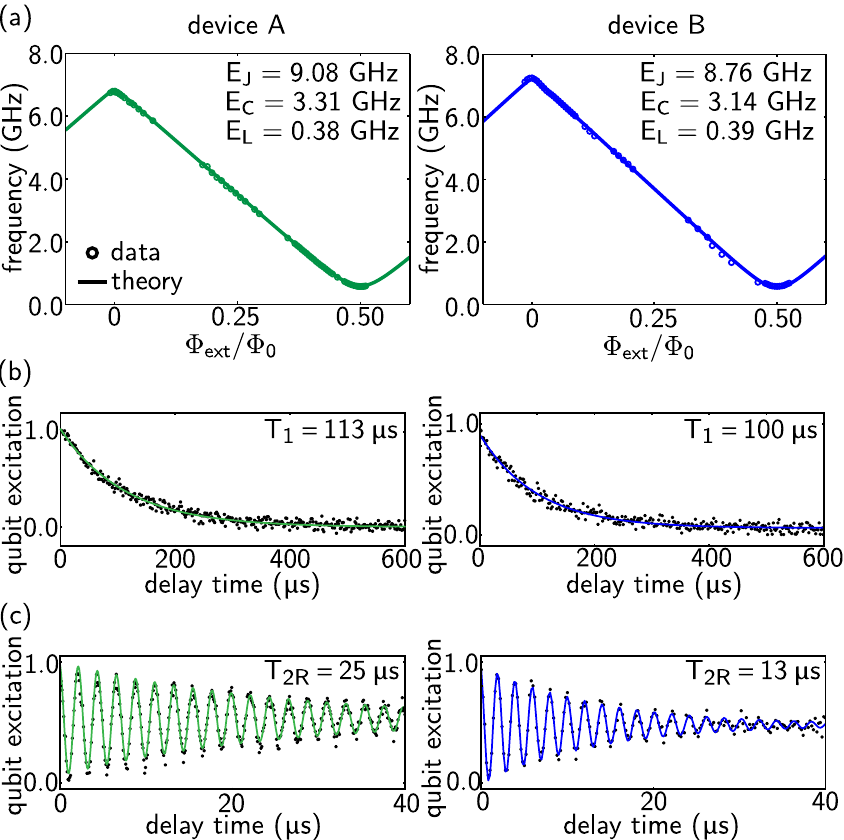}
\caption{(a) Ground-excited state transition frequency as a function of external flux for device A (green) and device B (blue). Two-tone spectroscopy data is shown in circles. Solid lines show theoretical fits obtained from numerical diagonalization using the indicated fit parameters. Relaxation times (b) and Ramsey coherence times (c) for device A and device B at $\fq{0.5}$.\label{fig:qubit_char}}
\end{figure}

\begin{figure*}
\includegraphics[width=172mm]{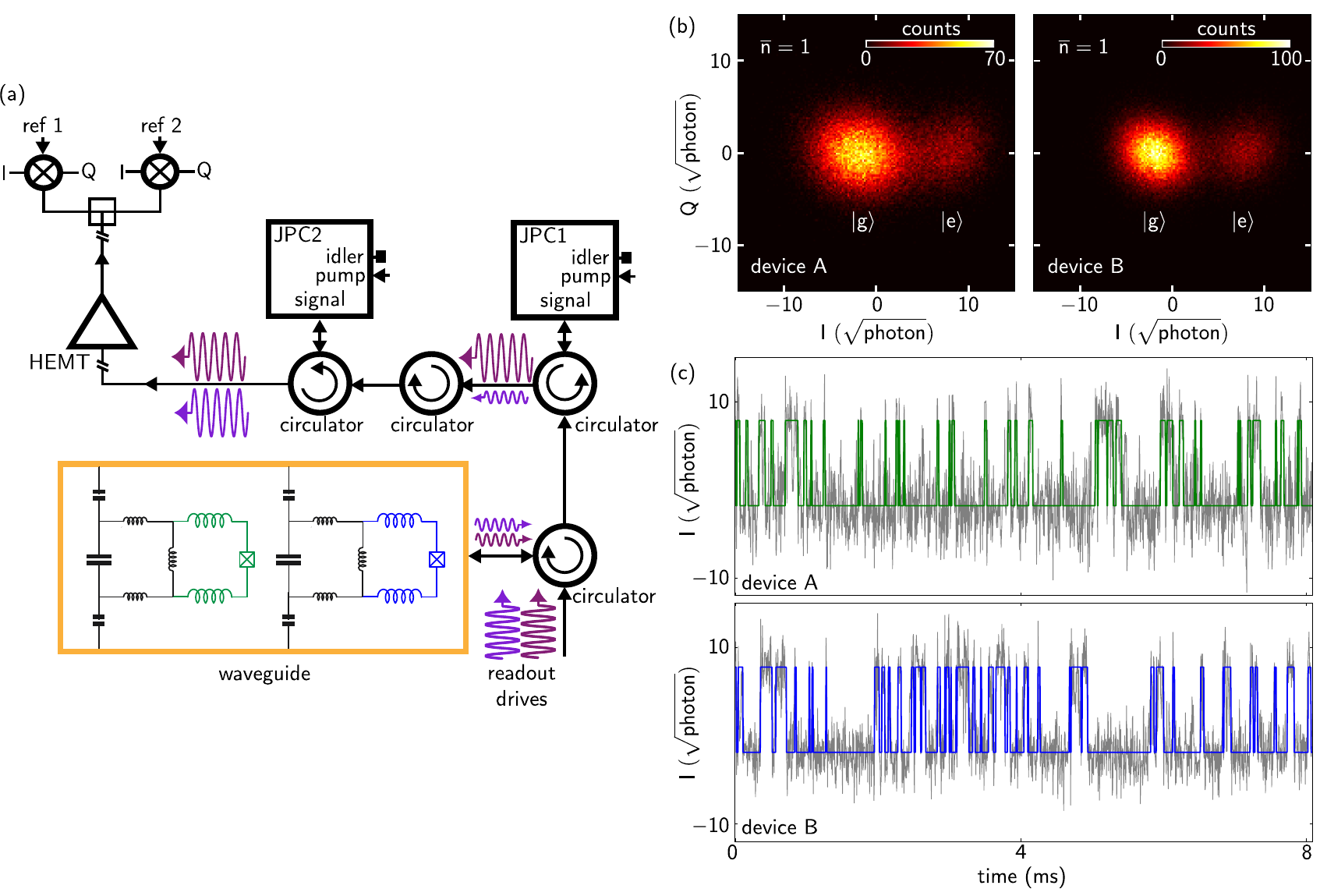}
\caption{(a) Measurement setup with cascaded Josephson parametric converters (JPCs) for simultaneous single-shot readout of quantum jumps. The output signals from the waveguide were preamplified in reflection by two JPCs connected in series. Each JPC was tuned to provide a gain of 20 dB at the resonant frequency of a readout antenna. The signals were subsequently amplified by a high electron-mobility transistor at 4 K. Finally, the signals were demodulated and digitized at room temperature using two heterodyne interferometer setups. (b) Histograms of simultaneously measured $I,Q$ quadratures in units of total number of photons per sample average for the fluxonium devices in equilibrium with their environment at $\fqn\approx0.5\Phi_0$. Each count corresponds to $5~\mu\text{s}$ of integration, and the total number of counts is 80,000. The state corresponding to each peak is labeled in white. The effective temperature was $\sim 20$~mK for device A and $\sim 25$~mK for device B. (c) Simultaneously measured quantum jump traces, which corresponded to the time evolutions of the $I$ quadratures of device A and device B. The raw traces are shown in grey and an estimate of the states of device A and device B are shown in green and blue, respectively.
 \label{fig:measurement}}
\end{figure*}

\section{Device Characterization}
Standard dispersive readout \cite{Blais2004} was used to measure the devices via microwave drives their respective readout antennae. We performed a two-tone spectroscopy experiment at different applied external magnetic flux ($\fqn$) points as shown in Fig.~\ref{fig:qubit_char}(a). The samples were biased via a large magnetic field coil wound around the waveguide that encircles both devices. We observed transitions between the ground and the excited states for both devices. Numerical diagonalization of the fluxonium Hamiltonian \cite{Smith2016} was used to fit the spectroscopic data shown with circles in Fig.~\ref{fig:qubit_char}(a). The fit parameters used to obtain the theoretical curves (solid lines in Fig.~\ref{fig:qubit_char}(a)) are indicated.

We confirmed the microwave hygiene of the electromagnetic environment by measuring the coherence times of the ground-excited state transition for device A and device B at $\fq{0.5}$. We refer to the ground and excited states of device A (B) as qubit A (B). Here, the qubit transition frequency for device A and device B were 565~MHz and 579~MHz, respectively. We performed standard time-domain measurements of the relaxation time ($T_1$) and Ramsey dephasing time ($T_{2R}$). The measured $T_1$'s for both qubits were in excess of 100$~\mu$s as shown in Fig.~\ref{fig:qubit_char}(b). The measured $T_{2R}$'s for qubit A and qubit B were 25~$\mu$s and 13~$\mu$s, respectively. The addition of an echo pulse into the standard Ramsey sequence did not extend the coherence times of the qubits, indicating that the coherence times of the two qubits are limited by noise characterized by time scales faster than several microseconds. 

\section{Simultaneous Measurements}

We demonstrate simultaneous monitoring of the fluxonium qubits with the following measurement setup. The output of the waveguide was fed into circulators, which then routed the output signals to two cascaded JPC quantum-limited amplifiers. The JPCs were tuned to provide a gain of 20 dB at $7.430~$GHz and $7.979~$GHz with bandwidths of 6~MHz and 5~MHz, respectively. Signals amplified in reflection by the JPCs were fed via circulators into a high electron-mobility transistor amplifier at 4~K. The amplified signals were then split at room temperature and demodulated at 50 MHz for device A and 25 MHz for device B using two heterodyne interferometer setups. A schematic of the full measurement setup is shown in Fig.~\ref{fig:measurement}(a).

Figure~\ref{fig:measurement}(b) shows the simultaneously measured $I$ and $Q$ quadratures of the two fluxonium qubits in equilibrium with their environment measured at $\fqn\approx0.5\Phi_0$ \cite{approxhalf}. The optimal measurement fidelity was achieved with a readout power corresponding to $\bar{n}= 1$ photon occupation of the readout resonator. Larger photon numbers resulted in faster measurements but also saturated the output of the JPCs as well as decreased the lifetimes of the two fluxonium qubits. This last effect has also been observed in transmons \cite{Boissonneault2008, Slichter2012}. We attained a readout fidelity of $95\%$ for each measurement with 5~$\mu$s of integration time. The total number of counts in each histogram is 80,000.

We observed the evolution of the $I$ quadratures of the two fluxonium qubits simultaneously, as shown in grey in Fig.~\ref{fig:measurement}(c). An estimate of the qubit state was determined using a two-point filter, similar to that used in Ref.~\cite{Vool2014}. The filter declared a change in the qubit state if the quadrature value crossed a threshold set $\sigma/2$ away from the new state, where $\sigma$ was one standard deviation away from the center of the peak corresponding to the new state obtained from the histogram shown in Fig.~\ref{fig:measurement}(b). Otherwise, the qubit was declared to remain in its previous state. The estimated qubit state is shown in green (blue) for device A (B) in Fig.~\ref{fig:measurement}(c).

\begin{figure}[t]
\includegraphics[width=86mm]{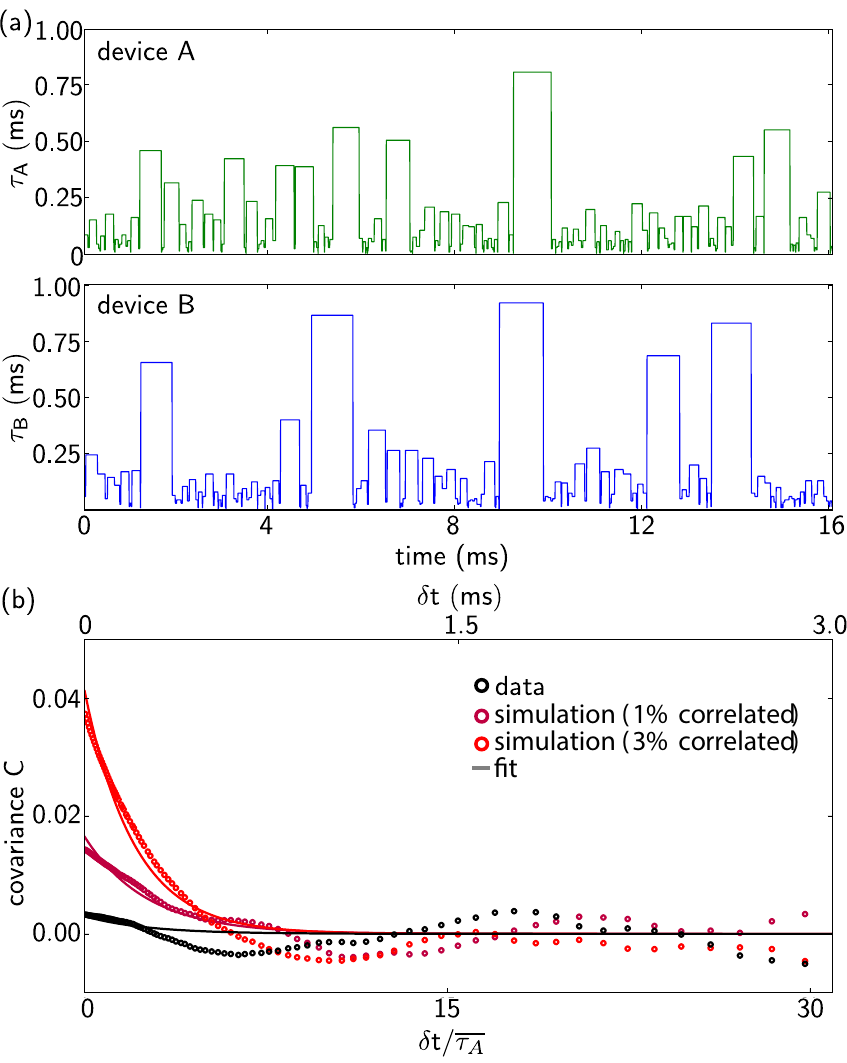}
\caption{(a) Time ($\tau_{A,B}$) each qubit spent in either $\g$ or $\e$ as a function of time (b) Covariance $C$ between $\tau_A$ at time t and $\tau_B$ at time $t + \delta t$ as a function of $\delta t$. The experimental data is indicated in black. Colored circles denote simulated data where the two devices simultaneously have the same $\tau$ for $1\%$ and $3\% $ of the time. Solid lines indicate exponential fits to the covariance. \label{fig:qubit_corr}}
\end{figure}

A qubit subject to frequent measurements of its energy stochastically jumps between its energy eigenstates. We therefore do not expect the states of the qubits to be correlated and indeed do not observe any correlations between the qubit states \cite{supplement}.The characteristic time scale over which a qubit changes its state is the $T_1$ of the qubit, which is determined by the relaxation channels coupled to the qubit. If the relaxation channels become coupled more strongly to the qubit or decrease in quality factor, the qubit will change its state more rapidly (i.e. its $T_1$ will decrease). Hence, to look for correlated relaxation channels, one should investigate the correlations between the times that the two qubits spend in either $\g$ or $\e$. Using the qubit state evolutions shown in Fig.~\ref{fig:measurement}(c), we extracted the total amount of time each qubit spends in a single state before a quantum jump occurs at each time step, which we denote as $\tau$. The evolutions of $\tau$ for device A and device B are shown in Fig.~\ref{fig:qubit_corr}(a). 

We examined the correlations present between $\tau_A$ and $\tau_B$ using the normalized covariance,
\begin{equation}
\begin{split}
C(t,t+\delta t)=\overline{\frac{\tau_A(t)\tau_B(t+\delta t)}{\overline{\tau_A}~\overline{\tau_B}}}-1,
\end{split}
\end{equation}
where $\tau_A(t)$ ($\tau_B(t)$) is the time that device A (B) spends in each state at time $t$, $\overline{\tau_A}~({\overline{\tau_B}})$ is the mean of each dataset, and $\delta t$ is the separation in time between the data taken for device A and the data taken for device B. The average for $C$ is taken over 2000 datasets of 20.48 ms of continuous monitoring.

In Fig.~\ref{fig:qubit_corr}(b), we also plot $C$ for simuled qubits with correlated jump times at $\delta t=0$ for different percentages of the total monitoring time. For correlated $\tau$ between device A and device B at time $\delta t=0$, we expect the covariance to decay on a time scale of the order of the mean time that the qubits spend in a state during the correlated times, and the amplitude to depend on the percentage of the total monitoring time that $\tau_A$ and $\tau_B$ are correlated. Simple exponential fits to $C$ are indicated by solid lines in Fig.~\ref{fig:qubit_corr}(b). The covariance of the measured devices corresponds to devices with correlated $\tau$ at $\delta t=0$ for $<0.5\%$ of the total monitoring time; the detection threshold of our experimental setup is $0.5\%$. We hence conclude that up to the detection efficiency of our experimental setup, the relaxation of the two devices is not correlated.


\section{Conclusion}

In conclusion, we have demonstrated simultaneous monitoring of fluxonium qubits using a low-loss waveguide. We used this architecture to investigate correlations between the relaxation times of the fluxonium qubits. We find no detectable correlations between their relaxation times. The detection efficiency could be improved by extending the lifetime of the fluxonium qubits, which may be achievable with a waveguide-based bandpass filter. 

In this experiment we have investigated correlations between relaxation mechanisms; correlations between dephasing mechanisms in different qubits can also be examined with this architecture when used in conjunction with the recently-demonstrated protocol for quantum non-demolition readout of the transverse component of a qubit \cite{Vool2016}. In such an experiment, correlations between the quantum jumps of the qubit measured in the $\sigma_x$-basis would be indicative of correlated dephasing mechanisms. In addition, this experiment can easily be extended to larger numbers of qubits by using quantum-limited amplifiers with higher bandwidths such as the traveling-wave parametric amplifier \cite{White2015,Macklin2015} or a tesselated three-wave mixing element \cite{Nick2016}. 

We have presented a general method for measuring correlations between the relaxation times of superconducting qubits. Our method can easily be incorporated into measurements made during quantum error correction. In future implementations of a quantum error correcting protocols, upon measurements of nonzero correlations, the experimentalist could change the applied error-correcting mechanism to target correlated errors and preserve the quantum information being processed. 

\section{Acknowledgements}

We acknowledge fruitful discussions with Rob Schoelkopf, Shyam Shankar, and Chen Wang. We thank Xu Xiao for his assistance with calculations. Facilities use was supported by YINQE, the Yale SEAS cleanroom, and NSF MRSEC DMR 1119826. This research was supported by ARO under Grant No. W911NF-14-1-0011 and by MURI-ONR Grant No. 0041302 (410042-3). 

\bibliographystyle{apsrevx}
\bibliography{simbib}

\begin{thebibliography}{31}%
\makeatletter
\providecommand \@ifxundefined [1]{%
 \@ifx{#1\undefined}
}%
\providecommand \@ifnum [1]{%
 \ifnum #1\expandafter \@firstoftwo
 \else \expandafter \@secondoftwo
 \fi
}%
\providecommand \@ifx [1]{%
 \ifx #1\expandafter \@firstoftwo
 \else \expandafter \@secondoftwo
 \fi
}%
\providecommand \natexlab [1]{#1}%
\providecommand \enquote  [1]{``#1''}%
\providecommand \bibnamefont  [1]{#1}%
\providecommand \bibfnamefont [1]{#1}%
\providecommand \citenamefont [1]{#1}%
\providecommand \href@noop [0]{\@secondoftwo}%
\providecommand \href [0]{\begingroup \@sanitize@url \@href}%
\providecommand \@href[1]{\@@startlink{#1}\@@href}%
\providecommand \@@href[1]{\endgroup#1\@@endlink}%
\providecommand \@sanitize@url [0]{\catcode `\\12\catcode `\$12\catcode
  `\&12\catcode `\#12\catcode `\^12\catcode `\_12\catcode `\%12\relax}%
\providecommand \@@startlink[1]{}%
\providecommand \@@endlink[0]{}%
\providecommand \url  [0]{\begingroup\@sanitize@url \@url }%
\providecommand \@url [1]{\endgroup\@href {#1}{\urlprefix }}%
\providecommand \urlprefix  [0]{URL }%
\providecommand \Eprint [0]{\href }%
\providecommand \doibase [0]{http://dx.doi.org/}%
\providecommand \selectlanguage [0]{\@gobble}%
\providecommand \bibinfo  [0]{\@secondoftwo}%
\providecommand \bibfield  [0]{\@secondoftwo}%
\providecommand \translation [1]{[#1]}%
\providecommand \BibitemOpen [0]{}%
\providecommand \bibitemStop [0]{}%
\providecommand \bibitemNoStop [0]{.\EOS\space}%
\providecommand \EOS [0]{\spacefactor3000\relax}%
\providecommand \BibitemShut  [1]{\csname bibitem#1\endcsname}%
\let\auto@bib@innerbib\@empty
\bibitem [{\citenamefont {Aharonov}\ and\ \citenamefont
  {Ben-Or}(1999)}]{Aharonov1999}%
  \BibitemOpen
  \bibfield  {author} {\bibinfo {author} {\bibfnamefont {D.}~\bibnamefont
  {Aharonov}}\ and\ \bibinfo {author} {\bibfnamefont {M.}~\bibnamefont
  {Ben-Or}},\ }\bibfield  {title} {\emph {\enquote {\bibinfo {title}
  {{Fault-Tolerant Quantum Computation With Constant Error Rate}},}\ }}\href
  {\doibase 10.1137/S0097539799359385} {\bibfield  {journal} {\bibinfo
  {journal} {Quantum}\ ,\ \bibinfo {pages} {63}} (\bibinfo {year}
  {1999})}\BibitemShut {NoStop}%
\bibitem [{\citenamefont {Aliferis}\ \emph {et~al.}(2005)\citenamefont
  {Aliferis}, \citenamefont {Gottesman},\ and\ \citenamefont
  {Preskill}}]{Aliferis2005}%
  \BibitemOpen
  \bibfield  {author} {\bibinfo {author} {\bibfnamefont {P.}~\bibnamefont
  {Aliferis}}, \bibinfo {author} {\bibfnamefont {D.}~\bibnamefont {Gottesman}},
  \ and\ \bibinfo {author} {\bibfnamefont {J.}~\bibnamefont {Preskill}},\
  }\bibfield  {title} {\emph {\enquote {\bibinfo {title} {{Quantum accuracy
  threshold for concatenated distance-3 codes}},}\ }}\href {\doibase
  http://www.rintonpress.com/xqic6/qic-6-2/097-165.pdf} {\bibfield  {journal}
  {\bibinfo  {journal} {Quant. Inf. Comp.}\ }\textbf {\bibinfo {volume} {6}},\
  \bibinfo {pages} {58} (\bibinfo {year} {2005})}\BibitemShut {NoStop}%
\bibitem [{\citenamefont {Aharonov}\ \emph {et~al.}(2006)\citenamefont
  {Aharonov}, \citenamefont {Kitaev},\ and\ \citenamefont
  {Preskill}}]{Aharonov2006}%
  \BibitemOpen
  \bibfield  {author} {\bibinfo {author} {\bibfnamefont {D.}~\bibnamefont
  {Aharonov}}, \bibinfo {author} {\bibfnamefont {A.}~\bibnamefont {Kitaev}}, \
  and\ \bibinfo {author} {\bibfnamefont {J.}~\bibnamefont {Preskill}},\
  }\bibfield  {title} {\emph {\enquote {\bibinfo {title} {{Fault-tolerant
  quantum computation with long-range correlated noise}},}\ }}\href {\doibase
  10.1103/PhysRevLett.96.050504} {\bibfield  {journal} {\bibinfo  {journal}
  {Phys. Rev. Lett.}\ }\textbf {\bibinfo {volume} {96}},\ \bibinfo {pages}
  {050504} (\bibinfo {year} {2006})}\BibitemShut {NoStop}%
\bibitem [{\citenamefont {Lidar}\ \emph {et~al.}(1998)\citenamefont {Lidar},
  \citenamefont {Chuang},\ and\ \citenamefont {Whaley}}]{Lidar1998}%
  \BibitemOpen
  \bibfield  {author} {\bibinfo {author} {\bibfnamefont {D.~A.}\ \bibnamefont
  {Lidar}}, \bibinfo {author} {\bibfnamefont {I.~L.}\ \bibnamefont {Chuang}}, \
  and\ \bibinfo {author} {\bibfnamefont {K.~B.}\ \bibnamefont {Whaley}},\
  }\bibfield  {title} {\emph {\enquote {\bibinfo {title} {{Decoherence-Free
  Subspaces for Quantum Computation}},}\ }}\href {\doibase
  10.1103/PhysRevLett.81.2594} {\bibfield  {journal} {\bibinfo  {journal}
  {Phys. Rev. Lett.}\ }\textbf {\bibinfo {volume} {81}},\ \bibinfo {pages}
  {2594} (\bibinfo {year} {1998})}\BibitemShut {NoStop}%
\bibitem [{\citenamefont {Viola}\ \emph {et~al.}(1999)\citenamefont {Viola},
  \citenamefont {Knill},\ and\ \citenamefont {Lloyd}}]{Viola1999}%
  \BibitemOpen
  \bibfield  {author} {\bibinfo {author} {\bibfnamefont {L.}~\bibnamefont
  {Viola}}, \bibinfo {author} {\bibfnamefont {E.}~\bibnamefont {Knill}}, \ and\
  \bibinfo {author} {\bibfnamefont {S.}~\bibnamefont {Lloyd}},\ }\bibfield
  {title} {\emph {\enquote {\bibinfo {title} {{Dynamical Decoupling of Open
  Quantum Systems}},}\ }}\href {\doibase 10.1103/PhysRevLett.82.2417}
  {\bibfield  {journal} {\bibinfo  {journal} {Phys. Rev. Lett.}\ }\textbf
  {\bibinfo {volume} {82}},\ \bibinfo {pages} {2417} (\bibinfo {year}
  {1999})}\BibitemShut {NoStop}%
\bibitem [{\citenamefont {Novais}\ and\ \citenamefont
  {Baranger}(2006)}]{Novais2006}%
  \BibitemOpen
  \bibfield  {author} {\bibinfo {author} {\bibfnamefont {E.}~\bibnamefont
  {Novais}}\ and\ \bibinfo {author} {\bibfnamefont {H.~U.}\ \bibnamefont
  {Baranger}},\ }\bibfield  {title} {\emph {\enquote {\bibinfo {title}
  {{Decoherence by correlated noise and quantum error correction}},}\ }}\href
  {\doibase 10.1103/PhysRevLett.97.040501} {\bibfield  {journal} {\bibinfo
  {journal} {Phys. Rev. Lett.}\ }\textbf {\bibinfo {volume} {97}},\ \bibinfo
  {pages} {040501} (\bibinfo {year} {2006})}\BibitemShut {NoStop}%
\bibitem [{\citenamefont {Paik}\ \emph {et~al.}(2011)\citenamefont {Paik},
  \citenamefont {Schuster}, \citenamefont {Bishop}, \citenamefont {Kirchmair},
  \citenamefont {Catelani}, \citenamefont {Sears}, \citenamefont {Johnson},
  \citenamefont {Reagor}, \citenamefont {Frunzio}, \citenamefont {Glazman},
  \citenamefont {Girvin}, \citenamefont {Devoret},\ and\ \citenamefont
  {Schoelkopf}}]{Paik2011}%
  \BibitemOpen
  \bibfield  {author} {\bibinfo {author} {\bibfnamefont {H.}~\bibnamefont
  {Paik}}, \bibinfo {author} {\bibfnamefont {D.~I.}\ \bibnamefont {Schuster}},
  \bibinfo {author} {\bibfnamefont {L.~S.}\ \bibnamefont {Bishop}}, \bibinfo
  {author} {\bibfnamefont {G.}~\bibnamefont {Kirchmair}}, \bibinfo {author}
  {\bibfnamefont {G.}~\bibnamefont {Catelani}}, \bibinfo {author}
  {\bibfnamefont {A.~P.}\ \bibnamefont {Sears}}, \bibinfo {author}
  {\bibfnamefont {B.~R.}\ \bibnamefont {Johnson}}, \bibinfo {author}
  {\bibfnamefont {M.~J.}\ \bibnamefont {Reagor}}, \bibinfo {author}
  {\bibfnamefont {L.}~\bibnamefont {Frunzio}}, \bibinfo {author} {\bibfnamefont
  {L.~I.}\ \bibnamefont {Glazman}}, \bibinfo {author} {\bibfnamefont {S.~M.}\
  \bibnamefont {Girvin}}, \bibinfo {author} {\bibfnamefont {M.~H.}\
  \bibnamefont {Devoret}}, \ and\ \bibinfo {author} {\bibfnamefont {R.~J.}\
  \bibnamefont {Schoelkopf}},\ }\bibfield  {title} {\emph {\enquote {\bibinfo
  {title} {{Observation of high coherence in Josephson junction qubits measured
  in a three-dimensional circuit QED architecture}},}\ }}\href {\doibase
  10.1103/PhysRevLett.107.240501} {\bibfield  {journal} {\bibinfo  {journal}
  {Phys. Rev. Lett.}\ }\textbf {\bibinfo {volume} {107}},\ \bibinfo {pages}
  {240501} (\bibinfo {year} {2011})}\BibitemShut {NoStop}%
\bibitem [{\citenamefont {Rigetti}\ \emph {et~al.}(2012)\citenamefont
  {Rigetti}, \citenamefont {Gambetta}, \citenamefont {Poletto}, \citenamefont
  {Plourde}, \citenamefont {Chow}, \citenamefont {C{\'{o}}rcoles},
  \citenamefont {Smolin}, \citenamefont {Merkel}, \citenamefont {Rozen},
  \citenamefont {Keefe}, \citenamefont {Rothwell}, \citenamefont {Ketchen},\
  and\ \citenamefont {Steffen}}]{Rigetti2012}%
  \BibitemOpen
  \bibfield  {author} {\bibinfo {author} {\bibfnamefont {C.}~\bibnamefont
  {Rigetti}}, \bibinfo {author} {\bibfnamefont {J.~M.}\ \bibnamefont
  {Gambetta}}, \bibinfo {author} {\bibfnamefont {S.}~\bibnamefont {Poletto}},
  \bibinfo {author} {\bibfnamefont {B.~L.~T.}\ \bibnamefont {Plourde}},
  \bibinfo {author} {\bibfnamefont {J.~M.}\ \bibnamefont {Chow}}, \bibinfo
  {author} {\bibfnamefont {A.~D.}\ \bibnamefont {C{\'{o}}rcoles}}, \bibinfo
  {author} {\bibfnamefont {J.~A.}\ \bibnamefont {Smolin}}, \bibinfo {author}
  {\bibfnamefont {S.~T.}\ \bibnamefont {Merkel}}, \bibinfo {author}
  {\bibfnamefont {J.~R.}\ \bibnamefont {Rozen}}, \bibinfo {author}
  {\bibfnamefont {G.~A.}\ \bibnamefont {Keefe}}, \bibinfo {author}
  {\bibfnamefont {M.~B.}\ \bibnamefont {Rothwell}}, \bibinfo {author}
  {\bibfnamefont {M.~B.}\ \bibnamefont {Ketchen}}, \ and\ \bibinfo {author}
  {\bibfnamefont {M.}~\bibnamefont {Steffen}},\ }\bibfield  {title} {\emph
  {\enquote {\bibinfo {title} {{Superconducting qubit in a waveguide cavity
  with a coherence time approaching 0.1 ms}},}\ }}\href {\doibase
  10.1103/PhysRevB.86.100506} {\bibfield  {journal} {\bibinfo  {journal} {Phys.
  Rev. B}\ }\textbf {\bibinfo {volume} {86}},\ \bibinfo {pages} {100506}
  (\bibinfo {year} {2012})}\BibitemShut {NoStop}%
\bibitem [{\citenamefont {Reed}\ \emph {et~al.}(2012)\citenamefont {Reed},
  \citenamefont {DiCarlo}, \citenamefont {Nigg}, \citenamefont {Sun},
  \citenamefont {Frunzio}, \citenamefont {Girvin},\ and\ \citenamefont
  {Schoelkopf}}]{Reed2012}%
  \BibitemOpen
  \bibfield  {author} {\bibinfo {author} {\bibfnamefont {M.~D.}\ \bibnamefont
  {Reed}}, \bibinfo {author} {\bibfnamefont {L.}~\bibnamefont {DiCarlo}},
  \bibinfo {author} {\bibfnamefont {S.~E.}\ \bibnamefont {Nigg}}, \bibinfo
  {author} {\bibfnamefont {L.}~\bibnamefont {Sun}}, \bibinfo {author}
  {\bibfnamefont {L.}~\bibnamefont {Frunzio}}, \bibinfo {author} {\bibfnamefont
  {S.~M.}\ \bibnamefont {Girvin}}, \ and\ \bibinfo {author} {\bibfnamefont
  {R.~J.}\ \bibnamefont {Schoelkopf}},\ }\bibfield  {title} {\emph {\enquote
  {\bibinfo {title} {{Realization of three-qubit quantum error correction with
  superconducting circuits.}}}\ }}\href {\doibase 10.1038/nature10786}
  {\bibfield  {journal} {\bibinfo  {journal} {Nature}\ }\textbf {\bibinfo
  {volume} {482}},\ \bibinfo {pages} {382} (\bibinfo {year}
  {2012})}\BibitemShut {NoStop}%
\bibitem [{\citenamefont {Kelly}\ \emph {et~al.}(2015)\citenamefont {Kelly},
  \citenamefont {Barends}, \citenamefont {Fowler}, \citenamefont {Megrant},
  \citenamefont {Jeffrey}, \citenamefont {White}, \citenamefont {Sank},
  \citenamefont {Mutus}, \citenamefont {Campbell}, \citenamefont {Chen},
  \citenamefont {Chen}, \citenamefont {Chiaro}, \citenamefont {Dunsworth},
  \citenamefont {Hoi}, \citenamefont {Neill}, \citenamefont {{O 'malley}},
  \citenamefont {Quintana}, \citenamefont {Roushan}, \citenamefont
  {Vainsencher}, \citenamefont {Wenner}, \citenamefont {Cleland},\ and\
  \citenamefont {Martinis}}]{Kelly2015}%
  \BibitemOpen
  \bibfield  {author} {\bibinfo {author} {\bibfnamefont {J.}~\bibnamefont
  {Kelly}}, \bibinfo {author} {\bibfnamefont {R.}~\bibnamefont {Barends}},
  \bibinfo {author} {\bibfnamefont {A.~G.}\ \bibnamefont {Fowler}}, \bibinfo
  {author} {\bibfnamefont {A.}~\bibnamefont {Megrant}}, \bibinfo {author}
  {\bibfnamefont {E.}~\bibnamefont {Jeffrey}}, \bibinfo {author} {\bibfnamefont
  {T.~C.}\ \bibnamefont {White}}, \bibinfo {author} {\bibfnamefont
  {D.}~\bibnamefont {Sank}}, \bibinfo {author} {\bibfnamefont {J.~Y.}\
  \bibnamefont {Mutus}}, \bibinfo {author} {\bibfnamefont {B.}~\bibnamefont
  {Campbell}}, \bibinfo {author} {\bibfnamefont {Y.}~\bibnamefont {Chen}},
  \bibinfo {author} {\bibfnamefont {Z.}~\bibnamefont {Chen}}, \bibinfo {author}
  {\bibfnamefont {B.}~\bibnamefont {Chiaro}}, \bibinfo {author} {\bibfnamefont
  {A.}~\bibnamefont {Dunsworth}}, \bibinfo {author} {\bibfnamefont {I.-C.}\
  \bibnamefont {Hoi}}, \bibinfo {author} {\bibfnamefont {C.}~\bibnamefont
  {Neill}}, \bibinfo {author} {\bibfnamefont {P.~J.~J.}\ \bibnamefont {{O
  'malley}}}, \bibinfo {author} {\bibfnamefont {C.}~\bibnamefont {Quintana}},
  \bibinfo {author} {\bibfnamefont {P.}~\bibnamefont {Roushan}}, \bibinfo
  {author} {\bibfnamefont {A.}~\bibnamefont {Vainsencher}}, \bibinfo {author}
  {\bibfnamefont {J.}~\bibnamefont {Wenner}}, \bibinfo {author} {\bibfnamefont
  {A.~N.}\ \bibnamefont {Cleland}}, \ and\ \bibinfo {author} {\bibfnamefont
  {J.~M.}\ \bibnamefont {Martinis}},\ }\bibfield  {title} {\emph {\enquote
  {\bibinfo {title} {{State preservation by repetitive error detection in a
  superconducting quantum circuit}},}\ }}\href {\doibase 10.1038/nature14270}
  {\bibfield  {journal} {\bibinfo  {journal} {Nature}\ }\textbf {\bibinfo
  {volume} {519}},\ \bibinfo {pages} {66} (\bibinfo {year} {2015})}\BibitemShut
  {NoStop}%
\bibitem [{\citenamefont {Rist{\`{e}}}\ \emph {et~al.}(2015)\citenamefont
  {Rist{\`{e}}}, \citenamefont {Poletto}, \citenamefont {Huang}, \citenamefont
  {Bruno}, \citenamefont {Vesterinen}, \citenamefont {Saira},\ and\
  \citenamefont {DiCarlo}}]{Riste2015}%
  \BibitemOpen
  \bibfield  {author} {\bibinfo {author} {\bibfnamefont {D.}~\bibnamefont
  {Rist{\`{e}}}}, \bibinfo {author} {\bibfnamefont {S.}~\bibnamefont
  {Poletto}}, \bibinfo {author} {\bibfnamefont {M.-Z.}\ \bibnamefont {Huang}},
  \bibinfo {author} {\bibfnamefont {A.}~\bibnamefont {Bruno}}, \bibinfo
  {author} {\bibfnamefont {V.}~\bibnamefont {Vesterinen}}, \bibinfo {author}
  {\bibfnamefont {O.-P.}\ \bibnamefont {Saira}}, \ and\ \bibinfo {author}
  {\bibfnamefont {L.}~\bibnamefont {DiCarlo}},\ }\bibfield  {title} {\emph
  {\enquote {\bibinfo {title} {{Detecting bit-flip errors in a logical qubit
  using stabilizer measurements}},}\ }}\href {\doibase 10.1038/ncomms7983}
  {\bibfield  {journal} {\bibinfo  {journal} {Nat. Commun.}\ }\textbf {\bibinfo
  {volume} {6}},\ \bibinfo {pages} {6983} (\bibinfo {year} {2015})}\BibitemShut
  {NoStop}%
\bibitem [{\citenamefont {Ofek}\ \emph {et~al.}(2016)\citenamefont {Ofek},
  \citenamefont {Petrenko}, \citenamefont {Heeres}, \citenamefont {Reinhold},
  \citenamefont {Leghtas}, \citenamefont {Vlastakis}, \citenamefont {Liu},
  \citenamefont {Frunzio}, \citenamefont {Girvin}, \citenamefont {Jiang},
  \citenamefont {Mirrahimi}, \citenamefont {Devoret},\ and\ \citenamefont
  {Schoelkopf}}]{Ofek2016}%
  \BibitemOpen
  \bibfield  {author} {\bibinfo {author} {\bibfnamefont {N.}~\bibnamefont
  {Ofek}}, \bibinfo {author} {\bibfnamefont {A.}~\bibnamefont {Petrenko}},
  \bibinfo {author} {\bibfnamefont {R.}~\bibnamefont {Heeres}}, \bibinfo
  {author} {\bibfnamefont {P.}~\bibnamefont {Reinhold}}, \bibinfo {author}
  {\bibfnamefont {Z.}~\bibnamefont {Leghtas}}, \bibinfo {author} {\bibfnamefont
  {B.}~\bibnamefont {Vlastakis}}, \bibinfo {author} {\bibfnamefont
  {Y.}~\bibnamefont {Liu}}, \bibinfo {author} {\bibfnamefont {L.}~\bibnamefont
  {Frunzio}}, \bibinfo {author} {\bibfnamefont {S.~M.}\ \bibnamefont {Girvin}},
  \bibinfo {author} {\bibfnamefont {L.}~\bibnamefont {Jiang}}, \bibinfo
  {author} {\bibfnamefont {M.}~\bibnamefont {Mirrahimi}}, \bibinfo {author}
  {\bibfnamefont {M.~H.}\ \bibnamefont {Devoret}}, \ and\ \bibinfo {author}
  {\bibfnamefont {R.~J.}\ \bibnamefont {Schoelkopf}},\ }\bibfield  {title}
  {\emph {\enquote {\bibinfo {title} {{Demonstrating quantum error correction
  that extends the lifetime of quantum information}},}\ }}\href {\doibase
  10.1038/nature18949} {\bibfield  {journal} {\bibinfo  {journal} {Nature}\
  }\textbf {\bibinfo {volume} {536}},\ \bibinfo {pages} {441} (\bibinfo {year}
  {2016})}\BibitemShut {NoStop}%
\bibitem [{\citenamefont {McDermott}\ \emph {et~al.}(2005)\citenamefont
  {McDermott}, \citenamefont {Simmonds}, \citenamefont {Steffen}, \citenamefont
  {Cooper}, \citenamefont {Cicak}, \citenamefont {Osborn}, \citenamefont {Oh},
  \citenamefont {Pappas},\ and\ \citenamefont {Martinis}}]{McDermott2005}%
  \BibitemOpen
  \bibfield  {author} {\bibinfo {author} {\bibfnamefont {R.}~\bibnamefont
  {McDermott}}, \bibinfo {author} {\bibfnamefont {R.~W.}\ \bibnamefont
  {Simmonds}}, \bibinfo {author} {\bibfnamefont {M.}~\bibnamefont {Steffen}},
  \bibinfo {author} {\bibfnamefont {K.~B.}\ \bibnamefont {Cooper}}, \bibinfo
  {author} {\bibfnamefont {K.}~\bibnamefont {Cicak}}, \bibinfo {author}
  {\bibfnamefont {K.~D.}\ \bibnamefont {Osborn}}, \bibinfo {author}
  {\bibfnamefont {S.}~\bibnamefont {Oh}}, \bibinfo {author} {\bibfnamefont
  {D.~P.}\ \bibnamefont {Pappas}}, \ and\ \bibinfo {author} {\bibfnamefont
  {J.~M.}\ \bibnamefont {Martinis}},\ }\bibfield  {title} {\emph {\enquote
  {\bibinfo {title} {{Simultaneous state measurement of coupled Josephson phase
  qubits.}}}\ }}\href {\doibase 10.1126/science.1107572} {\bibfield  {journal}
  {\bibinfo  {journal} {Science}\ }\textbf {\bibinfo {volume} {307}},\ \bibinfo
  {pages} {1299} (\bibinfo {year} {2005})}\BibitemShut {NoStop}%
\bibitem [{\citenamefont {Chen}\ \emph {et~al.}(2012)\citenamefont {Chen},
  \citenamefont {Sank}, \citenamefont {Omalley}, \citenamefont {White},
  \citenamefont {Barends}, \citenamefont {Chiaro}, \citenamefont {Kelly},
  \citenamefont {Lucero}, \citenamefont {Mariantoni}, \citenamefont {Megrant},
  \citenamefont {Neill}, \citenamefont {Vainsencher}, \citenamefont {Wenner},
  \citenamefont {Yin}, \citenamefont {Cleland},\ and\ \citenamefont
  {Martinis}}]{Chen2012}%
  \BibitemOpen
  \bibfield  {author} {\bibinfo {author} {\bibfnamefont {Y.}~\bibnamefont
  {Chen}}, \bibinfo {author} {\bibfnamefont {D.}~\bibnamefont {Sank}}, \bibinfo
  {author} {\bibfnamefont {P.}~\bibnamefont {Omalley}}, \bibinfo {author}
  {\bibfnamefont {T.}~\bibnamefont {White}}, \bibinfo {author} {\bibfnamefont
  {R.}~\bibnamefont {Barends}}, \bibinfo {author} {\bibfnamefont
  {B.}~\bibnamefont {Chiaro}}, \bibinfo {author} {\bibfnamefont
  {J.}~\bibnamefont {Kelly}}, \bibinfo {author} {\bibfnamefont
  {E.}~\bibnamefont {Lucero}}, \bibinfo {author} {\bibfnamefont
  {M.}~\bibnamefont {Mariantoni}}, \bibinfo {author} {\bibfnamefont
  {A.}~\bibnamefont {Megrant}}, \bibinfo {author} {\bibfnamefont
  {C.}~\bibnamefont {Neill}}, \bibinfo {author} {\bibfnamefont
  {A.}~\bibnamefont {Vainsencher}}, \bibinfo {author} {\bibfnamefont
  {J.}~\bibnamefont {Wenner}}, \bibinfo {author} {\bibfnamefont
  {Y.}~\bibnamefont {Yin}}, \bibinfo {author} {\bibfnamefont {A.~N.}\
  \bibnamefont {Cleland}}, \ and\ \bibinfo {author} {\bibfnamefont {J.~M.}\
  \bibnamefont {Martinis}},\ }\bibfield  {title} {\emph {\enquote {\bibinfo
  {title} {{Multiplexed dispersive readout of superconducting phase qubits}},}\
  }}\href {\doibase 10.1063/1.4764940} {\bibfield  {journal} {\bibinfo
  {journal} {Appl. Phys. Lett.}\ }\textbf {\bibinfo {volume} {101}},\ \bibinfo
  {pages} {182601} (\bibinfo {year} {2012})}\BibitemShut {NoStop}%
\bibitem [{\citenamefont {Schmitt}\ \emph {et~al.}(2014)\citenamefont
  {Schmitt}, \citenamefont {Zhou}, \citenamefont {Juliusson}, \citenamefont
  {Royer}, \citenamefont {Blais}, \citenamefont {Bertet}, \citenamefont
  {Vion},\ and\ \citenamefont {Esteve}}]{Schmitt2014}%
  \BibitemOpen
  \bibfield  {author} {\bibinfo {author} {\bibfnamefont {V.}~\bibnamefont
  {Schmitt}}, \bibinfo {author} {\bibfnamefont {X.}~\bibnamefont {Zhou}},
  \bibinfo {author} {\bibfnamefont {K.}~\bibnamefont {Juliusson}}, \bibinfo
  {author} {\bibfnamefont {B.}~\bibnamefont {Royer}}, \bibinfo {author}
  {\bibfnamefont {A.}~\bibnamefont {Blais}}, \bibinfo {author} {\bibfnamefont
  {P.}~\bibnamefont {Bertet}}, \bibinfo {author} {\bibfnamefont
  {D.}~\bibnamefont {Vion}}, \ and\ \bibinfo {author} {\bibfnamefont
  {D.}~\bibnamefont {Esteve}},\ }\bibfield  {title} {\emph {\enquote {\bibinfo
  {title} {{Multiplexed readout of transmon qubits with Josephson bifurcation
  amplifiers}},}\ }}\href {\doibase 10.1103/PhysRevA.90.062333} {\bibfield
  {journal} {\bibinfo  {journal} {Phys. Rev. A}\ }\textbf {\bibinfo {volume}
  {90}},\ \bibinfo {pages} {062333} (\bibinfo {year} {2014})}\BibitemShut
  {NoStop}%
\bibitem [{\citenamefont {Manucharyan}\ \emph {et~al.}(2009)\citenamefont
  {Manucharyan}, \citenamefont {Koch}, \citenamefont {Glazman},\ and\
  \citenamefont {Devoret}}]{Manucharyan2009}%
  \BibitemOpen
  \bibfield  {author} {\bibinfo {author} {\bibfnamefont {V.~E.}\ \bibnamefont
  {Manucharyan}}, \bibinfo {author} {\bibfnamefont {J.}~\bibnamefont {Koch}},
  \bibinfo {author} {\bibfnamefont {L.~I.}\ \bibnamefont {Glazman}}, \ and\
  \bibinfo {author} {\bibfnamefont {M.~H.}\ \bibnamefont {Devoret}},\
  }\bibfield  {title} {\emph {\enquote {\bibinfo {title} {{Fluxonium: Single
  Cooper-Pair Circuit Free of Charge Offsets}},}\ }}\href {\doibase
  10.1126/science.1175552} {\bibfield  {journal} {\bibinfo  {journal}
  {Science}\ }\textbf {\bibinfo {volume} {326}},\ \bibinfo {pages} {113}
  (\bibinfo {year} {2009})}\BibitemShut {NoStop}%
\bibitem [{\citenamefont {Pop}\ \emph {et~al.}(2014)\citenamefont {Pop},
  \citenamefont {Geerlings}, \citenamefont {Catelani}, \citenamefont
  {Schoelkopf}, \citenamefont {Glazman},\ and\ \citenamefont
  {Devoret}}]{Pop2014}%
  \BibitemOpen
  \bibfield  {author} {\bibinfo {author} {\bibfnamefont {I.~M.}\ \bibnamefont
  {Pop}}, \bibinfo {author} {\bibfnamefont {K.}~\bibnamefont {Geerlings}},
  \bibinfo {author} {\bibfnamefont {G.}~\bibnamefont {Catelani}}, \bibinfo
  {author} {\bibfnamefont {R.~J.}\ \bibnamefont {Schoelkopf}}, \bibinfo
  {author} {\bibfnamefont {L.~I.}\ \bibnamefont {Glazman}}, \ and\ \bibinfo
  {author} {\bibfnamefont {M.~H.}\ \bibnamefont {Devoret}},\ }\bibfield
  {title} {\emph {\enquote {\bibinfo {title} {{Coherent suppression of
  electromagnetic dissipation due to superconducting quasiparticles}},}\
  }}\href {\doibase 10.1083/nature13017} {\bibfield  {journal} {\bibinfo
  {journal} {Nature}\ }\textbf {\bibinfo {volume} {508}},\ \bibinfo {pages}
  {369} (\bibinfo {year} {2014})}\BibitemShut {NoStop}%
\bibitem [{\citenamefont {Vool}\ \emph {et~al.}(2014)\citenamefont {Vool},
  \citenamefont {Pop}, \citenamefont {Sliwa}, \citenamefont {Abdo},
  \citenamefont {Wang}, \citenamefont {Brecht}, \citenamefont {Gao},
  \citenamefont {Shankar}, \citenamefont {Hatridge}, \citenamefont {Catelani},
  \citenamefont {Mirrahimi}, \citenamefont {Frunzio}, \citenamefont
  {Schoelkopf}, \citenamefont {Glazman},\ and\ \citenamefont
  {Devoret}}]{Vool2014}%
  \BibitemOpen
  \bibfield  {author} {\bibinfo {author} {\bibfnamefont {U.}~\bibnamefont
  {Vool}}, \bibinfo {author} {\bibfnamefont {I.~M.}\ \bibnamefont {Pop}},
  \bibinfo {author} {\bibfnamefont {K.}~\bibnamefont {Sliwa}}, \bibinfo
  {author} {\bibfnamefont {B.}~\bibnamefont {Abdo}}, \bibinfo {author}
  {\bibfnamefont {C.}~\bibnamefont {Wang}}, \bibinfo {author} {\bibfnamefont
  {T.}~\bibnamefont {Brecht}}, \bibinfo {author} {\bibfnamefont {Y.~Y.}\
  \bibnamefont {Gao}}, \bibinfo {author} {\bibfnamefont {S.}~\bibnamefont
  {Shankar}}, \bibinfo {author} {\bibfnamefont {M.}~\bibnamefont {Hatridge}},
  \bibinfo {author} {\bibfnamefont {G.}~\bibnamefont {Catelani}}, \bibinfo
  {author} {\bibfnamefont {M.}~\bibnamefont {Mirrahimi}}, \bibinfo {author}
  {\bibfnamefont {L.}~\bibnamefont {Frunzio}}, \bibinfo {author} {\bibfnamefont
  {R.~J.}\ \bibnamefont {Schoelkopf}}, \bibinfo {author} {\bibfnamefont
  {L.~I.}\ \bibnamefont {Glazman}}, \ and\ \bibinfo {author} {\bibfnamefont
  {M.~H.}\ \bibnamefont {Devoret}},\ }\bibfield  {title} {\emph {\enquote
  {\bibinfo {title} {{Non-poissonian quantum jumps of a fluxonium qubit due to
  quasiparticle excitations}},}\ }}\href {\doibase
  10.1103/PhysRevLett.113.247001} {\bibfield  {journal} {\bibinfo  {journal}
  {Phys. Rev. Lett.}\ }\textbf {\bibinfo {volume} {113}},\ \bibinfo {pages}
  {247001} (\bibinfo {year} {2014})}\BibitemShut {NoStop}%
\bibitem [{\citenamefont {Lecocq}\ \emph {et~al.}(2011)\citenamefont {Lecocq},
  \citenamefont {Pop}, \citenamefont {Peng}, \citenamefont {Matei},
  \citenamefont {Crozes}, \citenamefont {Fournier}, \citenamefont {Naud},
  \citenamefont {Guichard},\ and\ \citenamefont {Buisson}}]{Lecocq2011}%
  \BibitemOpen
  \bibfield  {author} {\bibinfo {author} {\bibfnamefont {F.}~\bibnamefont
  {Lecocq}}, \bibinfo {author} {\bibfnamefont {I.~M.}\ \bibnamefont {Pop}},
  \bibinfo {author} {\bibfnamefont {Z.}~\bibnamefont {Peng}}, \bibinfo {author}
  {\bibfnamefont {I.}~\bibnamefont {Matei}}, \bibinfo {author} {\bibfnamefont
  {T.}~\bibnamefont {Crozes}}, \bibinfo {author} {\bibfnamefont
  {T.}~\bibnamefont {Fournier}}, \bibinfo {author} {\bibfnamefont
  {C.}~\bibnamefont {Naud}}, \bibinfo {author} {\bibfnamefont {W.}~\bibnamefont
  {Guichard}}, \ and\ \bibinfo {author} {\bibfnamefont {O.}~\bibnamefont
  {Buisson}},\ }\bibfield  {title} {\emph {\enquote {\bibinfo {title}
  {{Junction fabrication by shadow evaporation without a suspended bridge.}}}\
  }}\href {\doibase 10.1088/0957-4484/22/31/315302} {\bibfield  {journal}
  {\bibinfo  {journal} {Nanotechnology}\ }\textbf {\bibinfo {volume} {22}},\
  \bibinfo {pages} {315302} (\bibinfo {year} {2011})}\BibitemShut {NoStop}%
\bibitem [{SQU()}]{SQUIDnote}%
  \BibitemOpen
  \href@noop {} {\bibinfo  {journal} {The experiments in this paper were
  conducted at $\Phi_\text{ext}<0.5 \Phi_0$ in both fluxonium devices, which
  corresponded to $\Phi_\text{ext}< 0.01 \Phi_0$ through the superconducting
  interference devices in the antennae. We observed no change in the antennae
  frequencies over this range of $\Phi_\text{ext}$}\ }\BibitemShut {NoStop}%
\bibitem [{\citenamefont {Geerlings}(2013)}]{KurtisThesis}%
  \BibitemOpen
\bibfield  {journal} {  }\bibfield  {author} {\bibinfo {author} {\bibfnamefont
  {K.~L.}\ \bibnamefont {Geerlings}},\ }\href@noop {} {\emph {\bibinfo {title}
  {Improving Coherence of Superconducting Qubits and Resonators}}}\ (\bibinfo
  {publisher} {Yale University Press},\ \bibinfo {year} {2013})\BibitemShut
  {NoStop}%
\bibitem [{\citenamefont {Blais}\ \emph {et~al.}(2004)\citenamefont {Blais},
  \citenamefont {Huang}, \citenamefont {Wallraff}, \citenamefont {Girvin},\
  and\ \citenamefont {Schoelkopf}}]{Blais2004}%
  \BibitemOpen
  \bibfield  {author} {\bibinfo {author} {\bibfnamefont {A.}~\bibnamefont
  {Blais}}, \bibinfo {author} {\bibfnamefont {R.~S.}\ \bibnamefont {Huang}},
  \bibinfo {author} {\bibfnamefont {A.}~\bibnamefont {Wallraff}}, \bibinfo
  {author} {\bibfnamefont {S.~M.}\ \bibnamefont {Girvin}}, \ and\ \bibinfo
  {author} {\bibfnamefont {R.~J.}\ \bibnamefont {Schoelkopf}},\ }\bibfield
  {title} {\emph {\enquote {\bibinfo {title} {{Cavity quantum electrodynamics
  for superconducting electrical circuits: An architecture for quantum
  computation}},}\ }}\href {\doibase 10.1103/PhysRevA.69.062320} {\bibfield
  {journal} {\bibinfo  {journal} {Phys. Rev. A}\ }\textbf {\bibinfo {volume}
  {69}},\ \bibinfo {pages} {062320} (\bibinfo {year} {2004})}\BibitemShut
  {NoStop}%
\bibitem [{\citenamefont {Smith}\ \emph {et~al.}(2016)\citenamefont {Smith},
  \citenamefont {Kou}, \citenamefont {Vool}, \citenamefont {Pop}, \citenamefont
  {Frunzio}, \citenamefont {Schoelkopf},\ and\ \citenamefont
  {Devoret}}]{Smith2016}%
  \BibitemOpen
  \bibfield  {author} {\bibinfo {author} {\bibfnamefont {W.~C.}\ \bibnamefont
  {Smith}}, \bibinfo {author} {\bibfnamefont {A.}~\bibnamefont {Kou}}, \bibinfo
  {author} {\bibfnamefont {U.}~\bibnamefont {Vool}}, \bibinfo {author}
  {\bibfnamefont {I.~M.}\ \bibnamefont {Pop}}, \bibinfo {author} {\bibfnamefont
  {L.}~\bibnamefont {Frunzio}}, \bibinfo {author} {\bibfnamefont {R.~J.}\
  \bibnamefont {Schoelkopf}}, \ and\ \bibinfo {author} {\bibfnamefont {M.~H.}\
  \bibnamefont {Devoret}},\ }\bibfield  {title} {\emph {\enquote {\bibinfo
  {title} {{Quantization of inductively-shunted superconducting circuits}},}\
  }}\href {\doibase 10.1103/PhysRevB.94.144507} {\bibfield  {journal} {\bibinfo
   {journal} {Phys. Rev. B}\ }\textbf {\bibinfo {volume} {94}},\ \bibinfo
  {pages} {144507} (\bibinfo {year} {2016})}\BibitemShut {NoStop}%
\bibitem [{app()}]{approxhalf}%
  \BibitemOpen
  \href@noop {} {\bibinfo  {journal} {For simultaneous jump measurements,
  $\Phi_\text{ext}$ in device A was set to $0.5 \Phi_0$ and $\Phi_\text{ext}$
  in device B was set to $0.495 \Phi_0$}\ }\BibitemShut {NoStop}%
\bibitem [{\citenamefont {Boissonneault}\ \emph {et~al.}(2008)\citenamefont
  {Boissonneault}, \citenamefont {Gambetta},\ and\ \citenamefont
  {Blais}}]{Boissonneault2008}%
  \BibitemOpen
\bibfield  {journal} {  }\bibfield  {author} {\bibinfo {author} {\bibfnamefont
  {M.}~\bibnamefont {Boissonneault}}, \bibinfo {author} {\bibfnamefont {J.~M.}\
  \bibnamefont {Gambetta}}, \ and\ \bibinfo {author} {\bibfnamefont
  {A.}~\bibnamefont {Blais}},\ }\bibfield  {title} {\emph {\enquote {\bibinfo
  {title} {{Nonlinear dispersive regime of cavity QED: The dressed dephasing
  model}},}\ }}\href {\doibase 10.1103/PhysRevA.77.060305} {\bibfield
  {journal} {\bibinfo  {journal} {Phys. Rev. A}\ }\textbf {\bibinfo {volume}
  {77}},\ \bibinfo {pages} {060305} (\bibinfo {year} {2008})}\BibitemShut
  {NoStop}%
\bibitem [{\citenamefont {Slichter}\ \emph {et~al.}(2012)\citenamefont
  {Slichter}, \citenamefont {Vijay}, \citenamefont {Weber}, \citenamefont
  {Boutin}, \citenamefont {Boissonneault}, \citenamefont {Gambetta},
  \citenamefont {Blais},\ and\ \citenamefont {Siddiqi}}]{Slichter2012}%
  \BibitemOpen
  \bibfield  {author} {\bibinfo {author} {\bibfnamefont {D.~H.}\ \bibnamefont
  {Slichter}}, \bibinfo {author} {\bibfnamefont {R.}~\bibnamefont {Vijay}},
  \bibinfo {author} {\bibfnamefont {S.~J.}\ \bibnamefont {Weber}}, \bibinfo
  {author} {\bibfnamefont {S.}~\bibnamefont {Boutin}}, \bibinfo {author}
  {\bibfnamefont {M.}~\bibnamefont {Boissonneault}}, \bibinfo {author}
  {\bibfnamefont {J.~M.}\ \bibnamefont {Gambetta}}, \bibinfo {author}
  {\bibfnamefont {A.}~\bibnamefont {Blais}}, \ and\ \bibinfo {author}
  {\bibfnamefont {I.}~\bibnamefont {Siddiqi}},\ }\bibfield  {title} {\emph
  {\enquote {\bibinfo {title} {{Measurement-induced qubit state mixing in
  circuit QED from Up-converted dephasing noise}},}\ }}\href {\doibase
  10.1103/PhysRevLett.109.153601} {\bibfield  {journal} {\bibinfo  {journal}
  {Phys. Rev. Lett.}\ }\textbf {\bibinfo {volume} {109}},\ \bibinfo {pages}
  {153601} (\bibinfo {year} {2012})}\BibitemShut {NoStop}%
\bibitem [{sup()}]{supplement}%
  \BibitemOpen
  \href@noop {} {\bibinfo  {journal} {see Supplementary Information}\
  }\BibitemShut {NoStop}%
\bibitem [{\citenamefont {Vool}\ \emph {et~al.}(2016)\citenamefont {Vool},
  \citenamefont {Shankar}, \citenamefont {Mundhada}, \citenamefont {Ofek},
  \citenamefont {Narla}, \citenamefont {Sliwa}, \citenamefont {Zalys-Geller},
  \citenamefont {Liu}, \citenamefont {Frunzio}, \citenamefont {Schoelkopf},
  \citenamefont {Girvin},\ and\ \citenamefont {Devoret}}]{Vool2016}%
  \BibitemOpen
\bibfield  {journal} {  }\bibfield  {author} {\bibinfo {author} {\bibfnamefont
  {U.}~\bibnamefont {Vool}}, \bibinfo {author} {\bibfnamefont {S.}~\bibnamefont
  {Shankar}}, \bibinfo {author} {\bibfnamefont {S.~O.}\ \bibnamefont
  {Mundhada}}, \bibinfo {author} {\bibfnamefont {N.}~\bibnamefont {Ofek}},
  \bibinfo {author} {\bibfnamefont {A.}~\bibnamefont {Narla}}, \bibinfo
  {author} {\bibfnamefont {K.}~\bibnamefont {Sliwa}}, \bibinfo {author}
  {\bibfnamefont {E.}~\bibnamefont {Zalys-Geller}}, \bibinfo {author}
  {\bibfnamefont {Y.}~\bibnamefont {Liu}}, \bibinfo {author} {\bibfnamefont
  {L.}~\bibnamefont {Frunzio}}, \bibinfo {author} {\bibfnamefont {R.~J.}\
  \bibnamefont {Schoelkopf}}, \bibinfo {author} {\bibfnamefont {S.~M.}\
  \bibnamefont {Girvin}}, \ and\ \bibinfo {author} {\bibfnamefont {M.~H.}\
  \bibnamefont {Devoret}},\ }\bibfield  {title} {\emph {\enquote {\bibinfo
  {title} {{Continuous quantum nondemolition measurement of the transverse
  component of a qubit}},}\ }}\href {\doibase 10.1103/PhysRevLett.117.133601}
  {\bibfield  {journal} {\bibinfo  {journal} {Phys. Rev. Lett.}\ }\textbf
  {\bibinfo {volume} {117}},\ \bibinfo {pages} {133601} (\bibinfo {year}
  {2016})}\BibitemShut {NoStop}%
\bibitem [{\citenamefont {White}\ \emph {et~al.}(2015)\citenamefont {White},
  \citenamefont {Mutus}, \citenamefont {Hoi}, \citenamefont {Barends},
  \citenamefont {Campbell}, \citenamefont {Chen}, \citenamefont {Chen},
  \citenamefont {Chiaro}, \citenamefont {Dunsworth}, \citenamefont {Jeffrey},
  \citenamefont {Kelly}, \citenamefont {Megrant}, \citenamefont {Neill},
  \citenamefont {O'Malley}, \citenamefont {Roushan}, \citenamefont {Sank},
  \citenamefont {Vainsencher}, \citenamefont {Wenner}, \citenamefont
  {Chaudhuri}, \citenamefont {Gao},\ and\ \citenamefont
  {Martinis}}]{White2015}%
  \BibitemOpen
  \bibfield  {author} {\bibinfo {author} {\bibfnamefont {T.~C.}\ \bibnamefont
  {White}}, \bibinfo {author} {\bibfnamefont {J.~Y.}\ \bibnamefont {Mutus}},
  \bibinfo {author} {\bibfnamefont {I.-C.}\ \bibnamefont {Hoi}}, \bibinfo
  {author} {\bibfnamefont {R.}~\bibnamefont {Barends}}, \bibinfo {author}
  {\bibfnamefont {B.}~\bibnamefont {Campbell}}, \bibinfo {author}
  {\bibfnamefont {Y.}~\bibnamefont {Chen}}, \bibinfo {author} {\bibfnamefont
  {Z.}~\bibnamefont {Chen}}, \bibinfo {author} {\bibfnamefont {B.}~\bibnamefont
  {Chiaro}}, \bibinfo {author} {\bibfnamefont {A.}~\bibnamefont {Dunsworth}},
  \bibinfo {author} {\bibfnamefont {E.}~\bibnamefont {Jeffrey}}, \bibinfo
  {author} {\bibfnamefont {J.}~\bibnamefont {Kelly}}, \bibinfo {author}
  {\bibfnamefont {A.}~\bibnamefont {Megrant}}, \bibinfo {author} {\bibfnamefont
  {C.}~\bibnamefont {Neill}}, \bibinfo {author} {\bibfnamefont {P.~J.~J.}\
  \bibnamefont {O'Malley}}, \bibinfo {author} {\bibfnamefont {P.}~\bibnamefont
  {Roushan}}, \bibinfo {author} {\bibfnamefont {D.}~\bibnamefont {Sank}},
  \bibinfo {author} {\bibfnamefont {A.}~\bibnamefont {Vainsencher}}, \bibinfo
  {author} {\bibfnamefont {J.}~\bibnamefont {Wenner}}, \bibinfo {author}
  {\bibfnamefont {S.}~\bibnamefont {Chaudhuri}}, \bibinfo {author}
  {\bibfnamefont {J.}~\bibnamefont {Gao}}, \ and\ \bibinfo {author}
  {\bibfnamefont {J.~M.}\ \bibnamefont {Martinis}},\ }\bibfield  {title} {\emph
  {\enquote {\bibinfo {title} {{Traveling wave parametric amplifier with
  Josephson junctions using minimal resonator phase matching}},}\ }}\href
  {\doibase 10.1063/1.4922348} {\bibfield  {journal} {\bibinfo  {journal}
  {Appl. Phys. Lett.}\ }\textbf {\bibinfo {volume} {106}},\ \bibinfo {pages}
  {242601} (\bibinfo {year} {2015})}\BibitemShut {NoStop}%
\bibitem [{\citenamefont {Macklin}\ \emph {et~al.}(2015)\citenamefont
  {Macklin}, \citenamefont {O'Brien}, \citenamefont {Hover}, \citenamefont
  {Schwartz}, \citenamefont {Bolkhovsky}, \citenamefont {Zhang}, \citenamefont
  {Oliver},\ and\ \citenamefont {Siddiqi}}]{Macklin2015}%
  \BibitemOpen
  \bibfield  {author} {\bibinfo {author} {\bibfnamefont {C.}~\bibnamefont
  {Macklin}}, \bibinfo {author} {\bibfnamefont {K.}~\bibnamefont {O'Brien}},
  \bibinfo {author} {\bibfnamefont {D.}~\bibnamefont {Hover}}, \bibinfo
  {author} {\bibfnamefont {M.~E.}\ \bibnamefont {Schwartz}}, \bibinfo {author}
  {\bibfnamefont {V.}~\bibnamefont {Bolkhovsky}}, \bibinfo {author}
  {\bibfnamefont {X.}~\bibnamefont {Zhang}}, \bibinfo {author} {\bibfnamefont
  {W.~D.}\ \bibnamefont {Oliver}}, \ and\ \bibinfo {author} {\bibfnamefont
  {I.}~\bibnamefont {Siddiqi}},\ }\bibfield  {title} {\emph {\enquote {\bibinfo
  {title} {{A near-quantum-limited Josephson traveling-wave parametric
  amplifier}},}\ }}\href {\doibase 10.1126/science.aaa8525} {\bibfield
  {journal} {\bibinfo  {journal} {Science}\ }\textbf {\bibinfo {volume}
  {350}},\ \bibinfo {pages} {1} (\bibinfo {year} {2015})}\BibitemShut {NoStop}%
\bibitem [{\citenamefont {Frattini}\ \emph {et~al.}(2017)\citenamefont
  {Frattini}, \citenamefont {Vool}, \citenamefont {Shankar}, \citenamefont
  {Narla}, \citenamefont {Sliwa},\ and\ \citenamefont {Devoret}}]{Nick2016}%
  \BibitemOpen
  \bibfield  {author} {\bibinfo {author} {\bibfnamefont {N.~E.}\ \bibnamefont
  {Frattini}}, \bibinfo {author} {\bibfnamefont {U.}~\bibnamefont {Vool}},
  \bibinfo {author} {\bibfnamefont {S.}~\bibnamefont {Shankar}}, \bibinfo
  {author} {\bibfnamefont {A.}~\bibnamefont {Narla}}, \bibinfo {author}
  {\bibfnamefont {K.~M.}\ \bibnamefont {Sliwa}}, \ and\ \bibinfo {author}
  {\bibfnamefont {M.~H.}\ \bibnamefont {Devoret}},\ }\bibfield  {title} {\emph
  {\enquote {\bibinfo {title} {{3-Wave Mixing Josephson Dipole Element}},}\
  }}\href@noop {} {\bibfield  {journal} {\bibinfo  {journal}
  {arxiv:1702.00869}\ } (\bibinfo {year} {2017})}\BibitemShut {NoStop}%
\end{thebibliography}%

\end{document}